\documentstyle[12pt]{article}
\title{Dyons in  QCD: Confinement and Chiral Symmetry
Breaking\footnote{Lecture at the International School of Physics
"Enrico Fermi", Varenna, 27 June--7 July 1995}} \author{Yu.A.Simonov\\
Institute of Theoretical and Experimental Physics\\ 117259, Moscow,
B.Cheremushkinskaya 25, Russia} \date{} \newcommand{\be}{\begin{equation}}
\newcommand{\ee}{\end{equation}} 
\begin{document} \maketitle

\begin{abstract}
Dyonic classical solutions of
$SU(2)$ gluodynamics  are discussed. Exact form of dyonic solutions in
different gauges is presented and the nontrivial problem  of composition
of the dilute gas of dyons is settled.

Classical interaction between (anti)dyons is considered both analytically
and numerically. Confinement in the dyonic gas is discussed in connection
with the topological properties of individual dyon solution.

Fermionic zero modes of dyonic are displayed and the chiral symmetry
breaking in the dyonic gas is demonstrated.
\end{abstract}
\newpage

\section{Introduction}
The QCD vacuum is known to possess properties of confinement and
chiral symmetry breaking (CSB). In absence of dynamical quarks
(quenched approximation) confinement is characterized by the area law
of Wilson loop or zero average of Polyakov
line, while  CSB is connected to nonzero
values of chiral quark condensate. It was
found in lattice calculations  [1] that both
properties disappear at the same temperature
$T_c$, while a part of confining
configurations survive for $T>T_c$, ensuring
area law for spacial Wilson loops [2]
("magnetic confinement" [3]).

In addition, the QCD vacuum is characterized
by the topological susceptibility $\chi$
and nonperturbative energy density or gluonic
condensate [4], both quantities  imply (in
terms of gas of topological charges, like
instantons) a density of approximately 1 top
charge per $1 fm^4$ [5,6].

By now there is no model of QCD vacuum with
properties of confinement and CSB, based
directly on the QCD Lagrangian.

The most elaborated model is the instanton gas or liquid model (IM)
[6], which ensures CSB but lacks confinement [7]. Even so the
instanton model shows rather realistic features for hadron
correlators [8] demonstrating that CSB is already very important
property for correlators.

The same can be said about the Nambu--Jona-Lasinio model (NJL) [9]
(not directly connected to QCD Lagrangian) where confinement is also
absent.  Therefore in both  IM and NJL hadrons can    dissociate into
quarks.

Thus it is an urgent need to look for more realistic model of QCD
vacuum which obeys both basic properties: CSB and confinement. The
latter is associated widely with monopole-like degrees of freedom
[10], which may be of purely quantum or quasiclassical character. In
the latter case one should look for classical solutions of
Yang--Mills theory with monopole -- like fields. These solutions are
known for a long time [11]. Such solution can be obtained from the
multiinstanton solution in the so-called 'tHooft's ansatz
by the singular gauge tranformation [11]. In the general case of
finite--action multiinstanton solutions the form of fields
and generalized gauge transformation were found in [12].

The solutions have both color--electric and color--magnetic fields
and we therefore shall call them dyons.

The  dilute dyonic gas has been suggested some time ago as a model of
QCD vacuum  and some simple estimates of Wilson loop has been done
for dyons of finite time extension [12], demonstrating nonzero string
tension.  Recently the interest for the dyonic solutions has
revived.  In particular lattice studies of a classical and quantum
field of a dyon have been done and  a qualitative quasiabelian
 picture of confinement due to dyons was suggested [13].

Meanwhile the CSB properties of dyonic gas has been studied [14]. It
appears that each dyon has (infinitely many) zero fermionic modes [15]
and therefore CSB may occur through the same or similar mechanism as
in the IM, where each instanton has a zero fermionic mode and the gas
of instantons create collectively chiral quark mass and quark
condensate [16]. It was shown in [14] that this indeed happens, and
the values of chiral mass and condensates depend on density of dyons
in the gas.

Recently in an interesting series of papers [17] the properties of
the solvable $N=2$ SUSY, $4d$ model containing gluons, Majorana
fermions and an adjoint Higgs field were studied. It was demonstrated
[17] that there is a confining phase in the model with confinement
driven by the condensation of dyons and magnetic monopoles. With that
the dual-Meissner effect as confinement mechanism [10] obtains an
independent support, and also the dyonic gas model gets an additional
impetus.

It is a purpose of this lecture to consider general properties of
the dyon gas from point of view of confinement and CSB and to obtain
estimates of the corresponding parameters. To this end we refine our
ansatz of dyonic gas  done 10 years ago [12] and try to obtain a
self-consistent picture adjusting two parameters of the model:
average size of dyon $\rho$ and average distance between them, $R$.
With some choice of $\rho$ and $R$ we show that one can reasonably
reproduce values of string tension $\sigma$, quark condensate
$<\bar{q}q>$, gluonic condensate $G_2
=\frac{\alpha_s}{\pi}<(F_{\mu\nu}^a)^2>$, topological susceptibility $\chi$.
Moreover, one can calculate all field correlators and compare with existing
Monte-Carlo data.

As additional check of the model we study a possible scenario of temperature
phase transition and discuss nonperturbative physics at $T>T_c$.

The lecture is organized as follows. In chapter 2 we define a
single dyon solution, make a gas of dyons in chapter 3 and calculate
Wilson loop average for a single dyon and dyon gas in chapter 4. We
compare this result with the case of instanton and instantonic gas
and demonstrate the reason why dyons confine while instantons do not.
In computing string tension an important role is played by dyon-dyon
and dyon-antidyon correlations and screening length of dyon magnetic
and electric charge, which we specifically study.

In chapter 5 we calculate CSB for dyonic gas using resent results
[14], and expressing effect in terms of $\rho$ and $R$.

In conclusion the deconfinement scenario for dyonic gas is
discussed and some estimates for the dyonic gas are presented.

 \setcounter{equation}{0}
\renewcommand{\theequation}{2.\arabic{equation}}

\section{Properties of dyonic solutions}

We remind the classical Yang--Mills solution in the so--called
'tHooft ansatz
\be
A^a_{\mu}=-\frac{1}{g}\bar{\eta}^a_{\mu\nu}\partial_{\nu}lnW
\ee
where $\bar{\eta}^a_{\mu\nu}$ is 'tHooft symbol
\be
\bar{\eta}^a_{\mu\nu}= e_{a\mu\nu},~\mu,\nu=1,2,3~~  {\rm or} ~~
\delta_{a\nu}, ~\mu=4, {\rm or} -\delta_{a\mu}, ~\nu=4 \ee and \be
W=1+\sum^{N}_{i=1}\frac{\rho_i^2}{(x-x_i)^2}
\ee
$\rho_i$ and $x_i,~i=1,...~N$    are real; for finite $\rho_i=\rho$
one has the Harrington -- Shepard solutions [11], but we are
interested in the limit $\rho_i=\rho \to \infty$ and $x_i$ lying
equidistantly along the straight line \be x_k=\vec{r},~~kb,~~
k=0,~~\pm 1,~~\pm 2,~~... \pm N_1, \ee In most cases the limit of
$N_1\to \infty$ will be considered.

For the choice $\rho\to \infty$ one has
\be
W(r,t)=\sum_{k=0,\pm 1,...}\frac{1}{r^2+(t-kb)^2}
\ee
and $A^a_{\mu}$ due to (2.1) is in general periodic in time $t$. One can
make a gauge transformation
$\tilde{A}_{\mu}=\tilde{A}^a_{\mu}\frac{t_a}{2}=U^+
(A_{\mu}+\frac{i}{g}\partial_{\mu})U$,
 with $U=exp(i\frac{\vec{\tau}\vec{n}}{2}\theta)$, such that
 $\tilde{A}_{\mu}$ is time--independent for $N_1\to \infty$ [11,12].
 In this limit one has \be \tilde{A}_{ia}= f(r) e_{iba} n_b \ee \be
 \tilde{A}_{4a}= \varphi(r)  n_a
  \ee
 with
     \be
     f(r)=\frac{1}{gr}(1-\frac{\gamma r}{sh\gamma r})
     \ee
     \be
 \varphi(r)= \frac{1}{gr}(\gamma r cth \gamma r -1), ~~ \gamma
  =\frac{2\pi}{b}
   \ee
   We shall call (2.6)-(2.7) the Rossi solution [11] or the dyon
   solution since it has  both electric and magnetic fields; it is
   clear that Rossi solution belongs to the class of
   Prasad--Sommerfield solutions [11].

   One can easily find color--electric and color--magnetic  fields,
   $E_{ka}=-B_{ka}$
   \be
   B_{ka}=\delta_{ak}(-f'-f/r)+n_an_k(f'-f/r+gf^2)
   \ee
   The field (2.9)  contains both  long-range and short--range parts, indeed
   for large $r$ one has $f(r)\sim 1/gr$ and $B_{ka}\sim
   -\frac{n_an_k}{gr^2}$. To make separation more clear let us go from the
   hedgehog gauge(2.5-2.6) to the quasiabelian (or unitary) gauge, where the
   long--range part of $B_k$ is Abelian. This gauge rotation is given by the
   orthogonal matrix [18]
   \be
   0_{ik}=cos\theta \delta_{ik}+(1-cos \theta)\nu_i\nu_k +sin \theta
   e_{ikl}\nu_l
   \ee
   where we have introduced unit vector $\vec{e}$ and defined
   \be
   cos\theta \equiv \vec{e}\vec{n};~~\nu_i sin \theta =-e_{imn}e_mn_n
   \ee
   One also has property
   \be
   0_{ik}n_i=e_k
   \ee

   Therefore the gauge transformed $A'_{\mu a}$,
   \be
   A'_{\mu a}=\tilde{A}_{\mu b}\cdot 0_{ba}-\frac{1}{2g}
   0_{lb}\partial_{\mu}0_{lc}e{_abc}
   \ee
   yields
   \be
   A'_{4a}=\varphi(r) e_a
   \ee
   and
   \be
   B'_{ka} =(-2f/r+gf^2) n_ke_a +(-f'-f/r)[cos \theta
   \delta_{ka}+(1-cos \theta)\nu_k\nu_a-e_kn_a] \ee Now we notice
   that the long--range part -- the first term on the r.h.s.  of
   (2.16)  -- has a fixed color direction; by choosing $\vec{e}$
   along the third axis, we have \be -E'_k=B'_k(r\to \infty) \sim -
   \frac{1}{gr^2}\cdot n_k\cdot\frac{\tau_3}{2} \ee of course this
   transformation from (2.10) to (2.16) not defined
    at $\theta =\pi$. (Dirac string).
   We note also, that the long-range part of  $B_k$ and $E_k$ can be
   written as a derivate
     \be
   \vec{B}'=\nabla\Phi\frac{\tau_3}{2}+~\mbox{\rm short-range~ terms}
   \ee
   with
\be
   \Phi= \int^r_0(-2f/r'+gf^2) dr'
   \ee
   The total action of dyon is proportional to its time extension
   \be
   S=\frac{1}{2}\int d^3\vec{r}\int^T_0 dt (B^2_{ak}+E^2_{ak})=
   \frac{8\pi^2}{g^2b}T
   \ee
   It can be considered as a string of instantons of infinite radius [12],
   and for $N$  centers in  the  string (the "N-string") one has \be
    S(N)=\frac{8\pi^2}{g^2}(N-1)
   \ee
   The total  number of parameters for the case when we allow the centers to
   move and have finite radii $\rho_i$ is equal to $5N-1, (N>3)$ [19] (plus 3
   overall color orientations).

   Thus the enthropy of the $N$--string is smaller than that of $N$
   independent instantons -- in the latter case the total number of
   parameters is $8N$ due to independent color orientation of each
   instanton. This is the price one should pay for a new property of  dyon:
   its coherent field is able  to confine, as we shall demonstrate later.

   Now we turn to the quantum corrections  around the dyon. There is a
    literature  on the subject [20], but we shall use a simple method
    a l\'{a} Polyakov  to obtain the effective action of dyon
   due to quantum corrections around the dyonic classical solution.

   One can write the effective action $S_{eff} $ as [21]
   \be
   S_{eff}=\frac{1}{4}\int \frac{d^4q}{(2\pi)^4} (\frac{1}{g^2_0}+
   \pi(q^2))F^a_{\mu\nu}(q)F^a_{\mu\nu}(-q)
   \ee
   where the gluon plus ghost self--energy part $\pi(q^2)$ is
   \be
   \pi(q^2)= -\frac{11}{3} N_c\frac{1}{16\pi^2}ln \frac{\Lambda^2_0}{q^2}
   \ee
   and one can introduce the renormalized charge
   \be
   \frac{1}{g^2(q)}=\frac{1}{g^2_0}-\frac{11}{3}N_c\frac{1}{16\pi^2}ln
   \frac{\Lambda^2_0}{q^2}=\frac{b_0}{16\pi^2}ln \frac{q^2}{\Lambda^2}
   \ee

   For a long (anti)dyon $N\gg1)$ one can write
   \be
   \pm E_{ka} = B_{ka}=\zeta(q_0)b_{ka}(|\vec{q}|)
   \ee
   with $\zeta(q_0)=T\frac{sin(q_0T/2)}{(q_0T/2)}$. Introducing
   (2.24) in (2.21) one obtains
   \be
   S_{eff}\approx \frac{T}{b}\frac{8\pi^2}{g^2(\tilde{q})},
   \ee
   where $g^2(\tilde{q})$ is given in (2.24) and $\tilde{q}$ is
   obtained from the integral $\int
   b^2_{ka}\frac{(q)}{(2\pi)^3}\frac{d^3q}{g^2(q)} \approx
   \frac{1}{g^2(\tilde{q})}\frac{16\pi^2}{b}$
   \be
   \tilde{q}\approx \frac{2\gamma}{\pi}\approx \frac{4}{b}
   \ee
   The form (2.23) is valid when $q\gg\Lambda$; for small $q/\Lambda$
   one should replace (2.23) by the expression which takes into
   account confining configurations in the vacuum [21,22] -- dyonic
   gas in our case -- and one has approximately \be
   \frac{1}{g^2(q)}\approx \frac{b_0}{16\pi^2} ln
   \frac{q^2+m^2_0}{\Lambda^2},~~ m^2_0\approx 2\pi\sigma
   \ee

\setcounter{equation}{0}
\renewcommand{\theequation}{3.\arabic{equation}}

   \section{Dyonic gas}

    In this and following chapters the results of ref. [23]
    are largely used.
   Consider a system of several dyons and antidyons. We have several options
   for the composition of the system.

   (1) Dyonic gas as a simple superposition ansatz (similar to the instanton
   gas ansatz [6])
   \be
   A_{\mu}(x)=\sum^{N_+}_{i=1} A^{+(i)}_{\mu}(x)
   +\sum^{N_-}_{i=1}A_{\mu}^{-(i)}(x)
   \ee
   where $N_+$ and $N_-$ are numbers of dyons and antidyons respectively.
   Each individual vector potential $A^{(i)}_{\mu}$ (with superscript + for
   dyons and - for antidyons) can be characterized by a $4d$ vector
   $R^{(i)}$ and $0(4)$ unit vector $\omega^{(i)}$, defining direction of
   the straight line passing through $R^{(i)}$ and centers $kb$ in (2.5),
   moreover there is an overall color orientation $\Omega^{(i)}$, so that
      \be A^{(i)}_{\mu}(x)=\Omega^+_i(L\tilde{A})_{\mu}(r,t) \Omega_i \ee
      where $\tilde{A}_{\mu}(r,t)$ is the solution defined in
      chapter 2 and
      \be
      r=[(x-R^{(i)})^2-((x-R)\cdot \omega^{(i)}]^{1/2}~,
      \ee
      \be
      t=(x-R)\cdot  \omega^{(i)}~,
      \ee
      $L_{\mu\nu}(\omega)$ is the $0(4)$ rotation matrix, corresponding to
      $\omega$.

      Thus the overall vector potential $A_{\mu}(x)$ in (3.1) depends on
      the set\\ $\{\Omega_i, R^{(i)},
      \omega^{(i)}\},~i=1,...,N_++N_-$ and the total stochastic
      ensemble contains $A_{\mu}$ with all possible  values of this
      set.

      For the case of zero temperature, QCD vacuum should be $0(4)$
      invariant in the sense, that every observable $K$ is an
      average over stochastic ensemble of the operator $K(A)$ \be
      K=<K(A)>_{\Omega, R,\omega}
      \ee
      where the weight of averaging is Poincare--invariant with respect to
      $R, \omega$ and integration $d\Omega$ is with the usual Haar measure.

      For nonzero temperature there appears a preferred direction in $4d$.

      The relative simplicity of the ansatz (3.1) has a serious drawback
      --as also in the case of instanton gas -- $A_{\mu}(x)$ is not a
      classical solution. As a consequence even two dyons has an interaction
      energy (action), and there are some additional divergencies, which we
      now discuss.

      First of all one notices, that the action of the single dyon of fixed
      length $L$ is finite together with quantum corrections and
      proportional to $L$. This is because $B_{ka}=\pm E_{ka}\sim
      \frac{1}{r^2},~~r\to \infty$. For a pair of dyons situation is
      different. Let us consider the simplest case of $dd$  or $d\bar{d}$
      system with dyons in the same gauge (2.6-2.7), same
      $\omega^{(1)}=\omega^{(2)}$ -- two static dyons.

      \be
      A_{ia}=\frac{f(r_1)}{r_1}
      e_{iba}r_{1b}+\frac{f(r_2)}{r_2}e_{iba}r_{2b}
      \ee
      \be
      A_{4a}^{\pm}=\frac{\varphi(r_1)}{r_1}
      r_{1a}\pm \frac{\varphi(r_2)}{r_2}r_{2a}
      \ee
 and
$\vec{r}_i=\vec{r}-\vec{R}^{(i)},~~\vec{R}^{(i)}$ is position of a dyon.

For color magnetic and colorelectric field one has at large distances
\\ $r^{(i)}\gg \gamma^{-1}$
\be
 F_{12}=
\frac{h_1(\vec{r}_1\vec{\tau})}{2gr_1^4}+\frac{h_2(\vec{r}_2
\vec{\tau})}{2gr_2^4}- \frac{h_1\vec{\tau}\vec{r}_2
+h_2\vec{\tau}\vec{r}_1}
{2gr_1^2 r_2^2}
\ee
\be
E_{ia}=E_{ia}^{(1)}+E_{ia}^{(2)}+E_{ia}^{(12)}, h_i\equiv R_3^{(i)}
\ee
where $E_{ia}^{(1)}$ and $E_{ia}^{(2)} $ is given in (2.10)
and the interaction term $E_{ia}^{(12)}$ is
\be
E_{ia}^{(12)}=
\frac{\gamma}{g}[-\delta_{ia}\frac{(\vec{r}_1\vec{r}_2)}{r_1r_2}
(\frac{1}{r_2}\pm \frac{1}{r_1})+\frac{r_{1i}r_{2a}}{r_1r_2^2}\pm
\frac{r_{1a}r_{2i}}{r_1^2r_2}]
\ee
It is clearly seen in (3.10) that the colorelectric contribution to the
action
\be
S=\frac{1}{2}\int (E^2_{ia}+B^2_{ia})d^3\vec{r}dt
\ee
is diverging at large $r$ for the system of two dyons, since
$E_{ia}^{(12)}(dd)\sim 0(\frac{\gamma}{r}),~~ r\to \infty$; while for
$d\bar{d}$ system one has
\be
E_{ia}(d\bar{d})\sim 0(\frac{\gamma R}{r^2}),~~ r\to \infty~,~~
R=|\vec{R}^{(1)}-\vec{R}^{(2)}|
\ee
and the action is converging. From this it follows that it is impossible to
have thermodynamic limit for $N_+\neq N_-$ (as well as for the
Coulombic gas  with nonzero net charge, here situation is similar).

For the system of dyons and antidyons one can write for large distances
\be
B_{3a}=\sum^{N}_{i=1}B_{3a}^{(i)}-\sum^N_{i\not{=}j=1}
\frac{h_i r_{ja}}{2gr^2_ir^2_j}~,~~h_i\equiv R^{(i)}_3
\ee
\be
E_{ka}=\sum^{N}_{i=1}E_{ka}^{(i)}+
\frac{1}{g}\left\{
-\delta_{ka}\sum^N_{i\not{=}j=1}\frac{\vec{r}_i\vec{r}_j}{r_ir_j}
(\frac{\gamma_i}{r_j}+\frac{\gamma_j}{r_i})+\right.
\ee

$$
\left.\frac{r_{ik}r_{ja}\gamma_i}{r_ir_j^2}+r_{ia}
r_{jk}\frac{\gamma_j}{r_j^2r_j}
\right\}
$$
 where $\gamma_i=+\gamma
 $ and $-\gamma$ for dyons and antidyons respectively. One can see that the
 only dangerous term in (3.14) is $0(\frac{\gamma}{r})$ and it vanishes at
 large $r$ if
 $N_+=N_-$, so that $\sum^N_{i=1}\gamma_i=0$. Hence behaviour at
 large $r$ for $N_+=N_-$ is  $B_{ka}\sim
 E_{ka}\sim0(1/r^2)$.

 Let us now calculate action of $d\bar{d}$ system, eq. (3.11) as a
 function of dastance $R_{12}$ between $d$ and $\bar{d}$. The leading
 term comes from $(E^{(12)}_{ia})^2$ and has the form
 \be
 S_{int}\sim \frac{1}{2}\int (E_{ia}^{(12)})^2 d^3\vec{r}dt\sim
 T\gamma^2 R_{12}+0(T\gamma)+...
 \ee

 One observes in (3.15) linear confinement between $d$ and $\bar{d}$,
 $V_{int}=\frac{S_{int}}{t}\sim \gamma^2R_{12}$.

 Even more striking is the behaviour of the $d^2\bar{d}^2$ system.
 One can compute total action of the system and
 $S_{int}=S_{total}-\sum^4_{i=1} S_i$.

 $S_{int}$ depends on the configuration of the $d^2\bar{d}^2$ system
 when two $d$ and  two $\bar{d}$ are relatively close,  while
 distance between $d^2$ and $\bar{d}^2$, $R(d^2-\bar{d}^2)$, is large,
 one again obtains linear confinement
 \be
 S_{int}\sim T\gamma^2 R(d^2-\bar{d}^2)
 \ee
 For configuration of a "molecule" consisting of two atoms $d\bar{d}$
 with large $R(d\bar{d}-d\bar{d})$ one gets instead
 \be
 S_{int}\sim
 T\gamma^2\frac{(r(d\bar{d}))^4}{R^3(d\bar{d}-d\bar{d})},~~
 r(d\bar{d})\sim\gamma^{-1}
  \ee
 This behaviour reminds of a color Van-der-Waals potential between
 color--neutral objects.

 Thus a gas of static $(3d)$ dyons and antidyons with $N_+=N_-$
 degenerates  at minimal action to the gas of $d\bar{d}$ atoms of
 size $\gamma^{-1}$.  It is intersting to compute fields and  action
 of the $d\bar{d}$ system, when distance between $d\bar{d}$,
 $r(d\bar{d})$ is not large. At $r(d\bar{d})=0$ the colorelectric
 field vanishes, $E_{ia}=0$; while colormagnetic field $B_{ia}$
 becomes short range, $B_{ia}(r(d\bar{d})=0)\sim exp (-\gamma r),
 ~~r\to \infty$.
 The total action is
 \be
 S+T\frac{32\pi^2}{g^2b}\int^{\infty}_0\frac{dx}{sh^2x} [(1-xcth
 x)^2+2(1-x\cos echx)^2]
 \ee
 Numerical estimate yields
 \be
 S(r(d\bar{d})=0)\cong S_0\frac{T}{b} 1,1
 \ee
 This should be compared with the action of each (anti)dyon,
 $S=S_0\frac{T}{b}$.

 Thus one can see a strong attraction between $d$ and $\bar{d}$  at
 small distances.
 The overal behaviour of the $d\bar{d}$ action numerically computed
 in [23] is shown in Fig.1. One can see both regimes (3.19) and
 (3.15). However, the lack of long--distance field, and zero
 topological charge of the $d\bar{d}$ atom makes it useless from the
 point of confinement, as we shall see in the next Section.

 The situation with the dyonic--antidyonic gas in $4d$, i.e. when the
 (anti)dyon lines are have all directions, is different from the gas
 of static dyons. Indeed, let us take  one dyon with the line along
 4-th axis (i.e. static dyon) and antidyon with the line along the
 3d  axis. Assuming the superposition principle as in (3.6), one
 obtains for $E^{(12)}_{ia}$
 \be
 E^{(12)}_{ia}\sim \gamma^2e_{abc} n^{(1)}_b n_c^{(2)}
 \ee
 It is seen from (3.20) that the action (3.11) diverges for any
 distance between $d$ and $\bar{d}$, since $(E^{(12)}_{ia})^2$ is
 nonzero at large $r$.

 Thus the $4d$ gas of $d$ and $\bar{d}$ cannot exist in the ansatz
 (3.6-3.7)  and one must choose another composition principle,
  which we consider next.\\

 (2) \underline{The $4 d$ model of dyonic gas}

We are using again the superposition ansatz (3.1-3.4). At this point
one must specify in which gauge $A_{\mu}^{(i)}(x)$ in (3.1) are
summed up.

We have found only one gauge (modulo global rotations) where the
action of the superposition (3.1) is finite. This is actually the
singular gauge of the original 'tHooft ansatz (2.1),
$$
A_{\mu a}^{(i)}=-\bar{\eta}_{a\mu\nu}\partial_{\nu}\ln W^{(i)},
$$
\be
W^{(i)}=\frac{1}{2r}\frac{shr}{chr-cost}
\ee
yielding for a standard (not shifted and not rotated) solution the
form
$$
A_{ia}^{(i)}=e_{aik}n_k(\frac{1}{r}-cthr+\frac{shr}{chr-cost})-\delta_{ia}
\frac{sint}{chr-cost}
$$
\be
A_{4a}^{(i)}=n_a(\frac{1}{r}-cthr+\frac{shr}{chr-cost})
\ee
At large distances solutions (3.22) behave as
\be
A_{\mu
a}^{(i)}\sim\frac{1}{r},~~F_{\mu\nu}^{(i)}\sim\frac{1}{r^2}
\ee
and the same is true for the sum (3.1).

Hence the total action of a gas of dyons of finite length is finite.
The interaction energy of the $dd$ system, $V_{dd}$, and of the
$d\bar{d}$ system, $V_{d\bar{d}}$ in the gauge (3.22) depends on
distance $r$ and the relative time phase $\varphi$. At large $r$ both
$V_{dd}$ and $V_{d\bar{d}}$ fall off as $1/r$. This is shown in
Figs.2 and 3.
To ensure that dyonic gas could be a realistic model of the QCD
vacuum one must investigate the following points:

1) to check that the classical interaction between (anti) dyons is
weak enough at large distances, so that the dilute gas approximation
could be reasonably justified.

2) to prove the existence of the thermodynamic limit for the dyonic
ensemble (3.1), i.e. that the total action of the (big) volume $V_4$
is proportional to the volume, when it increases.

One must also prove that the free energy of the $d\bar{d}$ calculated
with quantum corrections has a minimum at a finite (and dilute)
density.

\setcounter{equation}{0}
\renewcommand{\theequation}{4.\arabic{equation}}

   \section{Confinement due to dyons}

We consider first the case of one dyon and calculate its contribution
to the Wilson loop in two ways:\\
 i) first we use the long range part
of the dyon field, appropriate for large Wilson loops as compared
with the dyon radius
ii)  we show that  Wilson loop can be rigorously computed through the
function $W$ of the 'tHooft ansatz and find that the (magnetic) flux
through the Wilson loop is $\pi$ for the dyon and $2\pi$ for
(multi)instanton; this  explains why confinement is present for the
first case and absent in the second in the simple picture of stochastic
confinement [24].

Finally we evaluate the Wilson loop for the $d\bar{d}$ gas.

i) Consider the circular Wilson loop as in Fig.4 and the dyon at the
distance $h$ above the  plane of the Wilson loop (to be the (12)
plane). We are using for a large loop of radius $R$ the quasiabelian
gauge form (2.17), which enables us to exploit the Stokes theorem
\be
W(C_R)=exp ig\int F_{12}d\sigma_{12}=exp(-i\frac{\tau_3}{2}\psi)
\ee
where $\psi$ is the solid angle for the geometry of Fig.4,
\be
\psi=2\pi h\int^R_0\frac{\rho
d\rho}{(\rho^2+h^2)^{3/2}}=2\pi(1-\frac{h}{\sqrt{h^2+R^2}})
\ee

One can see from (4.1)-(4.2) that for large $R\gg h$ (also $R\gg b$
 is necessary to use (2.17)) the color magnetic flux through the
 Wilson loop is
 \be
 flux(C)=\frac{\psi}{2}=\pi
 \ee
 As we shall discuss later, this is the condition for confinement in
 the dyonic gas, in the picture of stochastic confinement [24].

   ii) Consider now the general 'tHooft ansatz (2.1)-(2.3) and the
   same Wilson loop, Fig. 4, where for simplicity we put $h\ll R,
   h\to 0$. One has \be W(C_R)=exp~ig\int_{C_R} A_i dx_i= exp
   (\frac{i\tau_3}{2} 2\pi \frac{RW_r}{W})
   \ee
   where $W$ is given in (2.3), and $W_r$ is derivate of
   $W$ in $r=\sqrt{x^2}$ at $r=R$.

   Two cases are possible; a) for the "periodic instanton
   of Harrington--Shepard [11], of the size $\rho$, when $R\gg\rho$, one
   obtains
   \be
   W(C_R)\approx
exp(2 i\pi \tau_3\frac{\rho^2}{R^2}),
\ee
hence flux tends to zero (or $2\pi$) and no confinement results. The
same  is true for one instanton [25], in the opposite case,
$\rho\gg R$, again two possibilities appear, depending on the
total lengh of the instanton chain $L,~L=N_1b$.

Indeed, for $R\gg L$, one has
\be
\stackrel{\lim}{R\to\infty}
\frac{RW_r}{W}
=-2
 \ee
and the flux is $(-2\pi)$ - no confinement.

Finally for the infinite instanton chain, $L\to\infty$,
\be
W(R,x_4=0)=\frac{\gamma}{2R}\frac{sh \gamma R}{(ch \gamma R-1)}\to
\frac{\gamma}{2R}
\ee
and
\be
W(C_R)=\exp (-i\tau_3 \pi)= -1
\ee

This coincides with the result (4.1) for $h\ll R$. Thus the dyon
creates magnetic flux $(-\pi)$, while (multi)instanton creates
magnetic flux $(-2\pi)$. Consider now the two--dimensional gas of $d$
and $\bar{d}$ with $2d$ density $\frac{\bar{n}}{s}$ and with Poisson
distribution $w(n)$
\be
w(n)=e^{-\bar{n}}\frac{(\bar{n})^n}{n!}
\ee
The averaged Wilson loop can be written as
\be
<W(C_R)>=\sum_n e^{-\bar{n}}\frac{(-1)^n(\bar{n})^n}{n!}=
e^{-2\bar{n}}=e^{-\sigma S}
\ee
where the string tension $\sigma$ is : $\sigma=2\frac{\bar{n}}{S}$.

Thus the 2$d$ Poisson gas of dyons (and/or antidyons yields
confinement with string tension proportional to the ($2d$) average
density of dyons.

Next we consider the $3d$ gas dyons, which effectively means that in
the  total $4d$ ansatz (3.1) we keep for simplicity  all dyons with
roughly the same orientation $\omega^{(i)}\equiv (0,0,0,1)$. We again
can use the quasiabelian gauge (2.17) and write for the dilute gas
with $N$ dyons inside volume $V_3$
\be
<W(C_R)>=<exp(-\frac{i\tau_3}{2}\sum^N_{i=1}\psi^i)>=
<cos\frac{\bar{\psi}}{2}>^N,
\ee
where
\be
<cos\frac{\bar{\psi}}{2}>=\int\frac{d^3r}{V_3}
<cos\frac{{\psi}(r)}{2}.
\ee
Now for large volume $V_3,~V_3\gg R^3$, one has

\be
<W(C_R)>=exp(-\sigma S)
\ee
where we have defined
\be
\sigma=-\frac{N}{S} ln<cos
\frac{\bar{\psi}}{2}>
\ee

For  dyons distant from the Wilson loop one can write
\be
\psi(r)=\frac{\pi R^2}{r^2} cos \theta,
\ee
where $\theta$  is the angle beween the direction to the dyon (vector
$\vec{r}$) and the perpendicular to the Wilson loop plane, $R$ is
the radius of the Wilson loop. Expanding in (4.14) for small
$\psi^2\ll 1$ one has
\be
<\psi^2>\sim \frac{R^3}{V_3}; \sigma \approx C\frac{N}{V_3} R
\ee
In (4.16)
$c$ is a numerical constant, $N$--total number of dyons in the volume
$V_3$. Appearance of $R$ in (4.16) actually violates the area
law  eq. (4.13).  Indeed, $\sigma$ in (4.16) grows with the
radius of the Wilson loop indefinitely; this situation may be called
the \underline{superconfinement}.

One should note however, that the superconfinement occurs for the
ideal gas of $d$ and $\bar{d}$, when one neglects completely
correlations between $d$ and $\bar{d}$. In reality however for the
tightly correlated pair $d\bar{d}$ the long range field
disappears. Indeed,  adding  the vector potential (3.22) for the
dyon  and that of $\bar{d}$, which obtains by changing  the  sign
of $A^{(i)}_{4a}$ and the second term of $A^{(i)}_{ia}$ in (3.22)
one has
\be
A_{ia}(d\bar{d})=2e_{aik}n_k(\frac{1}{r}-coth
{}~r+\frac{sin~hr}{cos~hr-cost})
\ee
$$A_{4a}(d\bar{d})=0
$$
It is easy to calculate that the long range color magnetic and
color electric fields for vector potentials (4.17) disappear,
\be
E_{ia}(d\bar{d})=0,~~
B_{ia}(d\bar{d})=0(e^{-r})
\ee
Hence for a given dyon, antidyons can partly screen its field and
vice versa, and the phenomenon of Debye screening must take place.
One can estimate the Debye screening mass for the $4d$ gas of dyons
and antidyons [23]
\be
m^2_D\sim (\frac{N}{V_3})^{2/3},~~r_D=\frac{1}{m_D}
\ee
so that the Debye radius is of the order of average distance between
neighboring dyons. This means that dyons which are farther away from
the Wilson loop than $r_D$ do not participate in the creation of
string tension  and one should replace in (4.16) $R$ by $r_D$
for $R\gg r_D$. Finally one obtains an estimate for Debye
screened $d\bar{d}$ gas
\be
\sigma\approx const m^2 _D\sim (\frac{N}{V_3})^{2/3}
\ee
This behaviour of $\sigma$ ensures the area law for the Wilson
loop, the superconfinement due to the Debye screening transforms
into the  confinement.

These estimates have been checked in [23] by numerical calculations
of the Wilson loop.

In Figs. 5,6 the  contribution to the  Wilson  loop from a
dyon or a tight  $d\bar{d}$ pair is shown as a functions of
position of $d$ or $d\bar{d}$. One can see that the dyon
contribution is equal to $\pi$ when dyon is inside the
Wilson loop, and the $d\bar{d}$ pair contributes only when
it is exactly  on the Wilson contour.

\setcounter{equation}{0}
\renewcommand{\theequation}{5.\arabic{equation}}

\section{Chiral symmetry breaking in the dyonic gas}

We follow in this  chapter the recent  paper [14].
For the gas made of equal number of dyons $N_+$ and antidyons
$N_-,~~N_+=N_-=\frac{N}{2}$ in the big volume $V_4$ we assume that
the thermodynamic limit exists for the total action and other
extensive quantities like the free energy, when $N\to \infty,~~
V_4\to\infty$ and $\frac{N}{V_4}$ is fixed and finite.

We shall use for the dyonic gas the formalism  similar to that
exploited for the instanton gas by Diakonov and Petrov [5].
 To study the CSB as manifested in the nonzero chiral quark mass and chiral
condensate it is enough in case of instanton gas to consider
only the case of one flavour, $N_f=1$, since the so--called consistency
condition displaying CSB comes out the same also for $N_f=2,3$ [5,16].
Therefore we for simplicity confine ourselves in this section also to the
case $N_f=1$.

 The main driving mechanism for CSB is provided by the zero fermionic
modes on the topological charge [6,5]. For the instanton case zero
fermionic modes were found by 'tHooft [26], and later in [5] those have been
used to demonstrate CSB in the dilute instantonic gas.

In case of dyons fermionic modes have been  found in [27]. For our
purposes we consider two sets (They can be expressed one through another)
of zero modes, one with continuous parameter $\beta$ playing the role of
quasimomentum.
\be
\psi^{(\beta)}=W^{1/2}(\partial_0+i\partial_i\sigma_i)(W^{-1}F^{(\beta)})U_+
\ee
where $x_0\equiv x_4,$
\be
F^{(\beta)}=\sum^{\infty}_{n=-\infty}\frac{e^{i\beta n 2
\pi}}{r^2+(x_0+2\pi n)^2},
\ee
and $U_+$ is a constant  spinor of positive chirality.

Another set is labbeled by the integer $n$ and is obtained from
$\psi^{(\beta)}$ putting $\beta=0$ and keeping only one term in the sum
over $n$.
\be
u_n(x)= W^{1/2}(\partial_0+ i\partial_i\sigma_i)(W^{-1}
\frac{U_+}{r^2+(x_0+2\pi n)^2})
\ee
In what follows we shall use both sets.

The main problem to
be solved in this section is:  given fermionic zero modes on each of dyons
and antidyons; find the full quark Green's function for the dyonic gas with
the superposition ansatz (3.1).

To this end we make the same interpolating approximation for the
one-dyon quark Green's function $S^{(i)}$ as in [5], i.e. in the
exact spectral representation of $S^{(i)},~~i=1,~~...N$
\be
S^{(i)}(x,y)=\sum_n\frac{u_n^{(i)}(x)u^{(i)+}_n(y)}{\lambda_n-im}
\ee
containing all modes $n=1,2,...\infty,$
 we keep only zero modes $u^{(i)}_s$ and replace the nonzero-mode
 contribution by the free Green's function, since they coincide at
 large $n\sim \sqrt{p^2}$.

 Thus with $S_0=(-i\hat{D}(B)-im)^{-1}$ one has
 \be
 S^{(i)}(x,y)=S_0(x,y)+\sum_{{\rm
 zero~modes}}\frac{u_s^{(i)}(x)u_s^{(i)+}(y)}{-im}
 \ee

 One can see that $S^{(i)}$ diverges as $m\to 0$,  we shall show
 however that the total Green's function is finite for $m\to 0$ if
$ N_+=N_-$.

 Using (5.5) and (31) one derives the total Green's function
  to be
 \be
 S=S_0-\sum_{\stackrel{i,k}{n,m}}u_n^{(i)}(x)\left(
 \frac{1}{im+\hat{V}}\right)_{\stackrel{ik}{nm}}u^{(k)+}_m(y)
 \ee
 where upper indices $i,k$ run over all dyon numbers, $1\leq i,~k\leq
 N$, while lower indices $n,m$ run over all set of zero modes of the
 given dyon with the numbers $i,k$. We have also defined
 \be
 V^{ik}_{nm}\equiv \int
 u_n^{(i)+}(x)i(\hat{\partial}-ig\hat{B})u_m^{(k)}(x)d^4x
 \ee
 We keep here the field $B$ to make the formalism gauge invariant; in
 estimates we systematically put $B_{\mu}$ equal to zero. Note that
 $u^+_n$ and $u_m$ in (6.4) should have opposite chiralities, hence
 $V^{ik}$ refer to dyon--antidyon ($d\bar{d}$) or opposite
 ($\bar{d}d$) transitions, otherwise $V^{ik}$is zero.

 One can introduce graphs as in [5] to describe each term in (5.6)
 as a propagation amplitude from  a dyon $i$ to a dyon $k$ through
 scattering on many intermediate (anti)dyons centers, with scattering
 amplitude of each center (dyon) being $\frac{1}{im}$ and transition
 amplitude from center $j$ (excited to the s-th level) to center
 $l$ (excited to the r-th level) being $V^{jl}_{sr}$.

 The lower indices are not the only  new  element in (5.5)
 as compared to the instanton gas model [5,16]. The zero modes
 $u_n^{(i)}$ depend also on the Lorentz orientation $\omega^{(i)}$ of
 dyon, in addition to the color orientation $\Omega^{(i)}$ and
 position $R^{(i)}$ of the dyon, see Eq. (3.2).  \be
 u_n^{(i)}(x)=\Omega^{(i)}u_n^{(i)}(x-R^{(i)},\omega^{(i)})
 \ee
 Our next task is to compute the matrix elements of
 $(\frac{1}{im+\hat{V}})_{\stackrel{ik}{nm}}$ fixing initial and
 final states and averaging over all coordinates of intermediate
 dyons. To this end we introduce as in [5] the amplitudes
 $D^{ik}_{nm} $ and  $P^{ik}_{nm}$ for even and odd number of
 transitions $\hat{V}$ respectively
 \be
 (\frac{1}{im+\hat{V}})_{\stackrel{ik}{nm}}=
 \frac{\delta_{ik}\delta_{nm}}{im}+
 \left\{ \begin{array}{l}
 D_{nm}^{ik}
 (R_i^{(i)},R^{(k)},\Omega^{(i)},\Omega^{(k)},\omega^{(i)},\omega^{(k)})\\
 P_{nm}^{ik}
 (R_i^{(i)},R^{(k)},\Omega^{(i)},\Omega^{(k)},\omega^{(i)},\omega^{(k)})
 \end{array} \right.
 \ee
 In the definition (5.9) it is assumed that amplitudes of returns to
 the initial and final center $i$ ar $k$  are not included in
 $D^{ik},~P^{ik}$ and should be added separately (which makes
 Eq.(5.9) not an equality, but rather a symbolic equation). This
 amplitude of the return to the center $j$ we  denote as
 \be
 \Delta_{mn}= D^{jj}_{mn}
 (R^{(j)},R^{(j)},\Omega^{(j)},\Omega^{(j)},\omega^{(j)},\omega^{(j)})
 \ee
 Since in $D^{jj}$ integration over all intermediate coordinates
 $(R^{(k)},\Omega^{(k)},\omega^{(k)})$ is done, $\Delta_{mn}$ does
 not depend on
 $R^{(j)},\Omega^{(j)},\omega^{(j)}$ and  is a constant matrix.

  Taking into account any number of returns to the same center $j$,
 brings about a matrix $\varepsilon_{mn}$, defined as:
 \be
 \varepsilon_{mn}=\frac{1}{m}(1-im\hat{\Delta})_{mn}^{-1}
 \ee
 With its help the equations, connecting $\hat{P}$ and $\hat{D}$ can
 be written as follows
 \be
 P^{ik}_{nm}=-\frac{1}{im}V^{ik}_{nm}\frac{1}{im}-\frac{N}{2V_4}\int
 d^4R^{(j)}d\Omega^{(j)}d\omega^{(j)}
 \frac{1}{i}V^{ij}_{ns}\varepsilon_{sm'}D^{jk}_{m'm}
 \ee
 \be
 D^{ik}_{nm}=-\frac{N}{2V_4}\int
 d^4R^{(j)}d\Omega^{(j)}d\omega^{(j)}\frac{1}{i}V^{ij}_{ns'}
 \varepsilon_{s's}P^{jk}_{sm}
 \ee
  As a next step we separate out the dependence of
  $\hat{P},\hat{D},\hat{V}$ on lower indices and on
  $\Omega^{(i)},\Omega^{(k)}$. To this end we consider zero-mode
  solutions $u^{(i)}_n(x) $ in the form of (5.3) and make Fourier
  transform
  \be
  u^{(i)}_n(p)=\int u_n^{(i)}(x)e^{ipx}d^4x=e^{iP_02\pi
  n}\bar{u}^{(i)}(p)
  \ee
  It is important that $\bar{u}^{(i)}(p)$
  does not depend on $n$ altogether.  Therefore with the help of
  (5.7) one has
  \be
   V_{nm}^{ij} (R^{(i)},\Omega^{(i)},\omega^{(i)};
 R^{(j)},\Omega^{(j)},\omega^{(j)})
 =\int\frac{d^4p}{(2\pi)^4}e^{ip(R^{(i)}-
 R^{(j)})}v_{nm}^{ij}(p),
 \ee

  \be
  v_{nm}^{ij}(p) =e^{-2\pi i(p_0^i-p_0^jm)}\bar{u}^+(p^i)
\Omega^{+(i)}(-\hat{p})\Omega^{(j)}\bar{u}(p^j)
\ee
where $p^i=\Re_{\omega_i}p$, and $\Re_{\omega_i}$ is 0(4) rotation
 transforming time unit vector into $\omega_i$.

We introduce now "amputated" amplitudes $d,f,w$ instead of
$\hat{D},\hat{P}, \hat{V}$ as follows
\be
D^{ik}_{mn}=\int \frac{d^4p}{(2\pi)^4}
e^{ip(R^{(i)}-R^{(k)})-i2\pi(p^i_0m-p^k_0n)}\bar{u}^+(p^i)\Omega^{+(i)}
d(p^i,p^k)\Omega^{(k)}\bar{u}(p^k)
\ee
and similarly for $f(p^i,p^k);$ according to
(5.16)) one has $w(p^i,p^k)\equiv -\hat{p}$

Insertion of these definitions into Eqs.(5.12-5.13) yields
\be
f(p^i,p^k)=-\frac{w(p)}{(im)^2}-\frac{N}{2V_4N_c}\frac{w}{i}\int\nu
(p^j) d\omega^{(j)} d(p^j,p^k)
\ee
\be
d(p^i,p^k)=-\frac{N}{2V_4N_ci}w(p)\nu
(p^j) d\omega^{(j)} f(p^j,p^k)
\ee
where  we have introduced
\be
\nu(p)=\sum_{n,s} e^{+ip_02\pi n}\bar{u}(p)\varepsilon_{ns}
e^{-ip_02\pi s} \bar{u}^+(p)
\ee
One can see in (5.18-5.19) that $f$ and $d$ do not depend on
rotations in $p^i,p^k$ and the integration over $d\omega^i$ there
acts only on $\nu(p^j)$, so that with the definition
\be
\bar{\nu}(p)=\int\nu(p_j) d\omega^j
\ee
one
obtains
\be
d(p)=\frac{\frac{iN\hat{p}\bar{\nu}\hat{p}}{2V_4N_cm^2}}
{1+\left(\frac{N}{2V_4N_c}\right)^2\hat{p}\bar{\nu}\hat{p}\bar{\nu}}
\ee
and $f(p)$ is expressed through $d$ via (5.18). The definition (5.10)
can be used now to obtain the selfconsistency relation, taking into
account that at $m\to 0$, $\tilde{\Delta}\sim \frac{1}{m^2}$ and
therefore one has
\be
\Delta_{mn}\varepsilon_{ns}=\frac{i}{m^2}\delta_{ms}
\ee
as a result of insertion of (5.17) and (5.22) into (5.10) multiplied
with $\varepsilon_{mn}$, one has
\be
n_0=\frac{2V_4N_c}{N}\int
\frac{d^4p}{(2\pi)^4}\frac{M^2(p)}{M^2(p)+p^2}
\ee
where we have defined the average number of zero modes per dyon
$-n_0$, $n_0\approx \frac{V_4^{1/4}}{b},~b$ is the internal scale
parameter of dyons,

 We also introduced the chiral mass $M(p)$
 \be
 M(p)= \frac{N}{2V_4N_c} tr (\hat{p}\bar{\nu}(p)\hat{p})=
 \frac{N}{2V_4N_c}  p^2\bar{\nu}(p)
 \ee
 where we used the fact that $\bar{\nu}$ is averaged over all
 directions and should be proportional to the unit matrix in Lorentz
 and color space.

 Eq.(5.24) goes over into the corresponding consistency relation for
 instantons [5.16] when $n_0=1$ and matrix $\hat{\varepsilon}$ becomes a
 number, while $\bar{u}(p)$ is the Fourier transform of the 'tHooft's
 zero mode [26].

 The solution $d(p)$ (5.22) assumes the knowledge of the matrix
 $\varepsilon_{ns}$, while the consistency relation (5.24) imposes
 only one condition. Therefore the strategy of solution is as
 follows. From (5.17) one finds $\Delta_{mn}\equiv D^{ii}_{mn}=\int
 \frac{d^4p}{(2\pi)^4} e^{-2\pi ip_0(m-n)}\bar{u}^+(p)d(p)\bar{u}(p)$
 through $d(p)$. It clearly depends only on the modulus $|m-n|$. Then
 inverting the matrix $\Delta_{mn}$ one finds $\varepsilon_{mn}$ from
 (5.23). Finally from (5.20-5.21) one finds $\bar{\nu}(p)$ and
 inserts it into (5.22), defining $d(p)$ The cycle is thus
 completed, and should be repeated till  the convergence is achieved.

 One can also study another basis of zero modes, namely that of
 (5.1).

 In this case dependence on  $\beta$  can be also
 extracted, indeed
 $$
 u^{(i)}_{\beta}(p)=\sum_n e^{-i\beta 2\pi n} u^{(i)}_n=\sum_n
 e^{2\pi ni(p_0-\beta)}\bar{u}^{(i)}(p)=
 $$
 \be
 =\sum_k\delta(\beta-p_0-k)\bar{u}^{(i)}(p)
 \equiv\delta_{[\beta,p_0]}\bar{u}^{(i)}(p)
 \ee

 where we have introduced notation $\delta_{[\beta p_0]}$, implying
 that $\beta$ is in the interval [0,1] and $\delta$ -- function
 should be moderated first, introducing finite number of centers
 $N_0$ in the dyon $(\sum_{n=-N_0/2}^{N_0/2})$ and considering limit
 $N_0\to\infty$ at the end.

 In this way one obtains the same equations (5.17-5.19) for $f,d$ if
 the new definitions are used, e.g.
 \be
 D^{ik}_{\beta\beta'}=\int\frac{d^4p e^{ip(R^i-R^k)}}{(2\pi)^4}
\delta_{[\beta
 p_0]}\bar{u}^+(p^i)\Omega^{+i}d(p^ip^+)\Omega^k
 \bar{u}(p^k)\delta_{[\beta'p_0]}
 \ee
 and where in (5.18-5.19) now $\bar{\nu}$ is defined as
 \be
 \bar{\nu}(p)\to\tilde{\nu}(p)=\int d\omega
 \bar{u}(p)\varepsilon_{[p_0,p_0]}\bar{u}^+(p)
 \ee
 and
 \be
 \varepsilon_{[p_0,p_0]}\equiv \int^1_0 d\beta \int^1_0 d\beta'
 \delta_{[\beta,p_0]}\varepsilon_{\beta \beta'}\delta_{[\beta',p_0]}
 \ee
 From (5.27) one deduces  that
 $\Delta_{\beta\beta'}=\delta_{\beta\beta'}\Delta(\beta)$
  and hence also $\varepsilon_{\beta\beta'}$ is diagonal due to the
  relation
  \be
  \varepsilon_{\beta\beta'}\Delta_{\beta'\beta^{''}}
  =\frac{i}{m^2}\delta_{\beta\beta^{''}}
  \ee
  and is equal to
  \be
  \varepsilon_{\beta\beta'}=
  \delta_{\beta\beta^{'}}
  \varepsilon(\beta)=
  \delta_{\beta\beta^{'} }
  \frac{i}{m}\Delta^{-1}(\beta)
   \ee
  with
  \be
  \Delta(\beta)=\int\frac{d^4p}{(2\pi)^4}\delta_{[\beta
  p_0]}\bar{u}^+(p)d(p)\bar{u}(p)
  \ee
  The system (5.22), (5.24-5.29) is now
  complete.

  We now proceed to write down the quark propagator (5.6) in terms
  of functions $d, f$ and finally in terms of the chiral mass $M(p)$
  (5.25).

     Following the same procedure as in [5], one can rewrite (5.6)
     as
     $$
     S(p)=\frac{\hat{p}}{p^2}-\frac{N}{2V_4}\left(\frac{\delta_{ns}}{im}+
     \left(\Delta\frac{1}{1-im\Delta}\right)_{ns}\right)\times
     $$
     $$
     \times \int
     d\Omega^{(i)}d\omega^{(i)}(u_n^{(i)}(p,\omega^{(i)})u_s^{(i)+}
     (p,\omega^{(i)})+~{\rm dyon~}\leftrightarrow~ {\rm antidyon})-
     $$
     $$
     -\left(\frac{N}{2V_4}\right)^2\int d\Omega^{(i)} d\Omega^{(j)}
     d\omega^{(i)} d\omega^{(j)}
     u_n^{(i)}(p,\omega^{(i)}) (m\varepsilon_{ns})
     D^{ij}_{sl}(m\varepsilon_{lk})u^{+j}_k(p,\omega^j)
     $$
     $$
     -\left(\frac{N}{2V_4}\right)^2\int d\Omega^{(i)} d\Omega^{(j)}
     d\omega^{(i)} d\omega^{(j)}
     u_n^{(\bar{i})}(p,\omega^{(i)}) (m\varepsilon_{ns})
     D^{\bar{i}\bar{j}}_{sl}(m\varepsilon_{lk})u^{+\bar{j}}_k(p,\omega^j)
     $$
   $$
     -\left(\frac{N}{2V_4}\right)^2\int d\Omega^{(i)} d\Omega^{(j)}
     d\omega^{(i)} d\omega^{(j)}
     [u_{n}^{(i)}(p,\omega^{(i)}) (m)\varepsilon_{ns}
     P^{i\bar{j}}_{sl}(m)\varepsilon_{lk}u^{+\bar{j}}_k(p,\omega^j)
     $$
     \be
     +(i\to\bar{i},
     \bar{j}\to j)]
     \ee

      Submitting in
     (5.33) expression (5.17-5.19) and (5.22), (5.24)  we finally
     obtain $S(p)$ in the form
     \be
     S(p)=\frac{\hat{p}+iM(p)}{p^2+M^2}
     \ee
     This form justifies the meaning of $M(p)$ as  a chiral mass,
     i.e.  an effective mass of quark due to CSB. It coincides with
     the form of  $S(p)$ for the instantonn gas [5], however the
      explicit expression for $M(p)$ (5.25) differs.

     The most remarkable feature of (5.34) is the disappearance of
     the massless pole $\frac{\hat{p}}{p^2}$ from $S_0(p)$ One should
     have in mind of course that the form (5.34) is
     gauge--noninvariant and obtained neglecting confinement. If one
     takes into account these effects, as in [27], the pole structure
     in (5.34) is supplemented by the area law due to the string between
     the given quark and an antiquark and the pole is never present
     in physical amplitudes.

     From (5.34) one can easily compute the chiral condensate:
$$
     <\bar{q}q>_{Mink.}=-i<\bar{q}q>_{Eucl.}=
     i<tr S(x,x)>=
$$
\be
=   i\int\frac{d^4p}{(2\pi)^4} S(p)=
     -4N_c\int \frac{d^4p}{(2\pi)^4}\frac{M(p)}{p^2+M^2(p)}
     \ee
     It is nonzero thus confirming the phenomenon
     of CSB in the dilute dyonic gas.

\section{Conclusions  and prospectives}

We have given arguments that the dilute dyonic gas provides confinement and
CSB and therefore may be a good candidate for a realistic quasiclassical QCD
vacuum. Additional numerical checks are necessary of the area
law of the Wilson loop for the $0(4)$ invariant $4d$ dyonic
gas, which are now in progress [23]. If confirmed, the dyonic
gas ansatz can be used in the same program of detailed
calculations as were done for the instanton gas [8]. In
addition one can calculate field correlators and condencates to
be used as input in OPE and the vacuum correlator method
[3,22,27].

 Meanwhile in anticipation of exact numerical checks let us
 estimate roughly parameters of dyonic gas which could ensure
 realistic values of 1) gluonic condensate 2) chiral condensate
 3) string tension  4) topological susceptibility. To get 1)
 and 4) at realistic values one needs roughly density of one
 topological  charge per $1 fm^4$. This can be saturated by
 $3d$ density of dyons of 1 dyon per $1 fm^3$ and with
 $b\approx 1 fm$. Then the average size of dyon is
 $\gamma^{-1}=\frac{b}{2\pi}\approx 0.16 fm$ and one  expects
 from Eqs. (5.24), (5.25) and (5.35) to get a realistic (within
 a factor of 2-3) chiral condensate. Finally, with the given
 dyon density  the string tension (4.20) will be of a
 reasonable order of  magnitude, $\sigma\sim$ several units
 $\times fm^{-2}$. Thus order of magnitude estimates show that
 realistic model of dyonic gas is feasible.

 As a last point in this lecture we discuss now a possible
 scenario of temperature phase transition in the dyonic gas
 vacuum.

 The confined phase is described by the gas of dyons -- better to
 say, gas of dyonic lines which are oriented in all directions;
 at $T=0$ these directions $(\omega_{\mu}^{(i)}$, see Eq.
 (3.2)) are spread uninformly in the $0(4)$, but at $T>0$ the
 distribution of $\omega_{\mu}^{(i)} $ may be deformed. It is
 important, that $d$ and $\bar{d}$ are not paired, i.e. the
 correlation length of a $d\bar{d}$ pair is of the order of
 average distance between $D$ and $\bar{d}$, or $n^{-1/3}$.

 The deconfined phase can be chosen in such a way, that all
 dyons with lines directed along axis 1,2,3 are paired, i.e.
 $d$ and $\bar{d}$ form neutral $d\bar{d}$ atoms with average
 size of $\gamma^{-1}$; producing no long -- range field. Therefore
  string tension in the planes (14), (24) and (34)  vanishes
-- there is no confinement in the usual sense.

However, dyons and antidyons with lines $\omega_{\mu}^{(i)}$
along the 4-th axis are not paired, the $d\bar{d}$ average
distance is of the order of $n^{-1/3}$ and the confinement in
the spacial planes (1,2), (1,3) and (2,3) persists. One can
distinguish between two phases. If in the confining phase one
can write $$
S_{tot}^{conf}=\sum^4_{\nu=1}(S_{\nu}(d)+S_{\nu}(\bar{d}))=8S_0,~~
S_0=\frac{8\pi^2}{g^2}\frac{L}{b}
$$
then assuming that the number of $d$ and $\bar{d}$ does not
change, in the deconfined phase one obtains
$$
S_{tot}^{deconf}=\sum^3_{\nu=1}S_{\nu}(d\bar{d})+S_4(d)+S_4(\bar{d})\cong
5.3 S_0, $$
where we have put $S_{\nu}(d\bar{d})\approx 1.1. S_0$ in agreement with the
estimate (3.19). Thus one can see that the gluon condensate changes by some
40\% across the phase transition; this fact roughly agrees with the magnetic
confinement model of ref. [3], yielding reasonable
 estimates of $T_c$ (see lecture "Hot nonperturbative
 QCD" by the  same author).

  One can see that the dyonic gas model  may  explain
  the deconfinement transition in a sensible way,
  however exact  numerical computations are necessary to
  elaborate the detailed picture.

  This work was financially supported by the Russian Fund
  for Fundamental Research grant 95-02-05436.

\newpage
\begin{center}

{\bf  Figure captions}\\

\end{center}

Fig. 1. The interaction energy, $V(r)\equiv S_{int}/T$, for  a dyon
   and an antidyon in the static gauge, Eqs. (3.6-3.7) at the
   distance $r$ $vs$ $r/b$ From Ref.[23].\\

 Fig. 2. The interaction energy $V(r,\varphi)=<S_{int}>/T$, in units
 of $S_0/T$ where $S_{int}$ is averaged over the time period $b$, for
 the $dd$ system as a  function of distance
 $\Gamma_{dd}/b\equiv r$ and relative time phase
 $\varphi$. The logarithmic singularity at $r=0,
 \varphi=0$ is cut off by hand. From ref. [23].\\

 Fig. 3. The same as in Fig.2 for the $d\bar{d}$ system.
 The absolute minimum of $V_{d\bar{d}}$  is at
 $\varphi=\pi, r=0$ and is equal to $-1.3 S_0/T$.\\

 Fig. 4.  The Wilson loop of radius $R$ in the
 (1,2)--plane and  a dyon at the distance $h$ above the
 plane.\\

 Fig. 5. Contribution to the Wilson  loop from the dyon
 placed at distance $ h=zR$ above the plane of the loop
 and at distance $r.R$ from the center of the loop  $vs$
 $z$ and $r$. From ref. [23].\\

 Fig. 6. Contribution to the Wilson loop from  the tight $d\bar{d}$
 pair. Notations are the same as in Fig.4. From ref. [23].\\

 \end{document}